\documentclass[12pt]{article}%
\usepackage{citesort}
\usepackage{amsmath}
\usepackage{graphicx}%
\usepackage{amsfonts}%
\usepackage{amssymb}
\begin{document}

\begin{center}
\bigskip

\textbf{{\large Remark on double diffractive $\chi$ meson production}}
\end{center}

\vskip             3mm

\begin{center}
Adam Bzdak
\end{center}

\vskip             3mm

\begin{center}
M. Smoluchowski Institute of Physics, Jagiellonian University \\
Reymonta 4, 30-059 Krak\'{o}w, Poland

E-mail: \texttt{bzdak@th.if.uj.edu.pl}
\end{center}

\vskip                                   .5cm

\begin{quotation}
The double pomeron exchange contributions to the central inclusive and
exclusive $\chi_{c}^{0}$ and $\chi_{b}^{0}$ mesons production in the
Bialas-Landshoff approach are calculated. We find the model to be consistent
with the preliminary CDF upper limit on double diffractive exclusive $\chi
_{c}^{0}$ production cross section.

\bigskip

\end{quotation}

\section{Introduction}

The study of the double pomeron exchange (DPE) production processes is
interesting in its own right. It is an ideal way to improve our understanding
of diffractive processes and the dynamics of the pomeron exchange.

However, the great interest in such reactions is caused by the possibility of
the DPE processes to be one of the main mechanisms leading to Higgs boson
production
\cite{Schafer,Bial-Land,Cudell,Levin,Khoze-all,Cox,Peschan-PRL,Ingelman,Petrov,Bzdak-H}
within a very clean experimental environment.

In the present Letter we are particularly interested in the exclusive and
central inclusive (central inelastic) DPE production of heavy quarkonium
states $\chi_{c}$ and $\chi_{b}$ \cite{KMR-chi1,KMR-chi2,Feng}, see also
\cite{Stein,Peng}. In the exclusive DPE event the central object $\chi$ is
produced alone, separated from the outgoing hadrons by rapidity gaps:%
\[
p\bar{p}\rightarrow p+\text{gap}+\chi+\text{gap}+\bar{p}.
\]
In the central inclusive DPE event an additional radiation accompanying the
central object is allowed.

The basis for our considerations is the Bialas-Landshoff model for central
inclusive double diffractive Higgs boson production \cite{Bial-Land}. We
showed that the Bialas-Landshoff model and its exclusive extension
\cite{Bzdak-H} give satisfactory description of the DPE central inclusive and
exclusive dijet cross sections \cite{Bzdak-jj,Bzdak-H}.

In this Letter we show that the exclusive extension of the Bialas-Landshoff
model is consistent with the preliminary CDF upper limit on double diffractive
exclusive $\chi_{c}^{0}$ production cross section \cite{Koji}.

\section{Central inclusive $\chi$ meson production}

\begin{figure}[h]
\begin{center}
\includegraphics[width=3.80cm,height=2cm]{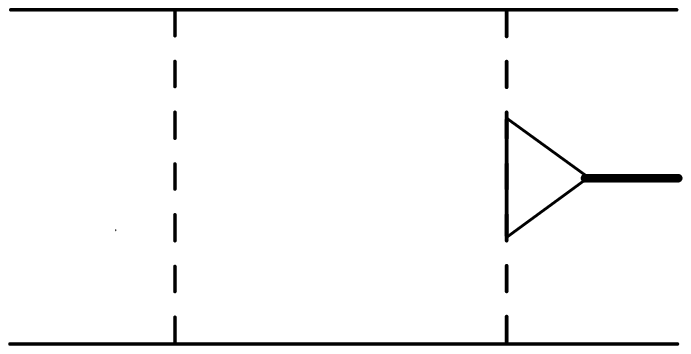}  \hspace{0.2cm}
\includegraphics[width=3.80cm,height=2cm]{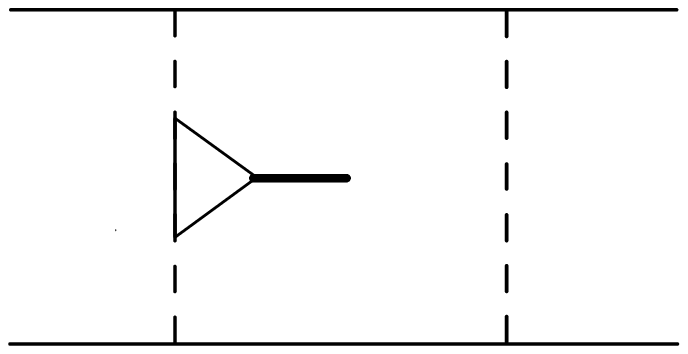}  \vskip         5mm
\includegraphics[width=3.80cm,height=2cm]{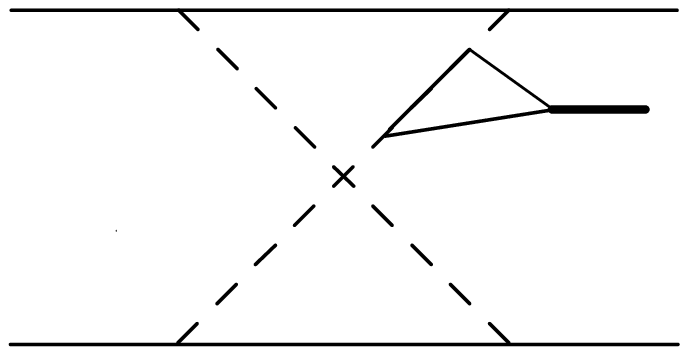}  \hspace{0.2cm}
\includegraphics[width=3.80cm,height=2cm]{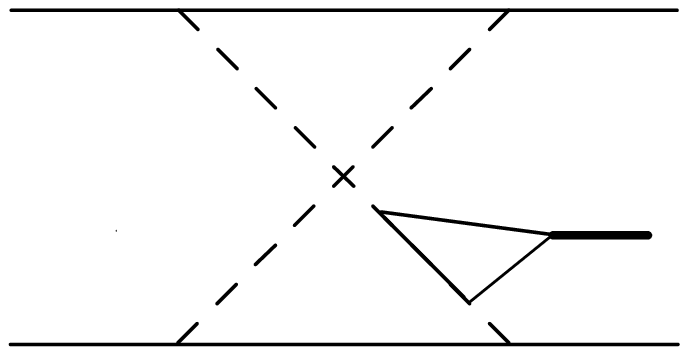}
\end{center}
\caption{Four diagrams contributing to the amplitude of the process of $\chi$
meson production by double pomeron exchange. The dashed lines represent the
exchange of the non-perturbative gluons. The $\chi$ coupling is taken to be
through a $c$-quark and $b$-quark loop for $\chi_{c}$ and $\chi_{b}$
respectively.}%
\label{chi-all}%
\end{figure}

In the Bialas-Landshoff approach pomeron exchange corresponds to the exchange
of a pair of non-perturbative gluons which takes place between a pair of
colliding quarks \cite{pomeron}. Thus, the process of $\chi$ meson production
is described by a sum of the four (reggeized) diagrams shown in Fig.
\ref{chi-all} where dashed lines correspond to the non-perturbative gluons.
The $\chi$ coupling is taken to be through a $c$-quark and $b$-quark loop for
$\chi_{c}$ and $\chi_{b}$ respectively.

Our calculation follows closely that of Ref. \cite{Bial-Land} where the DPE
contribution to the central inclusive Higgs boson production is calculated.

\begin{figure}[ptb]
\begin{center}
\includegraphics[width=5.30cm,height=4.20cm]{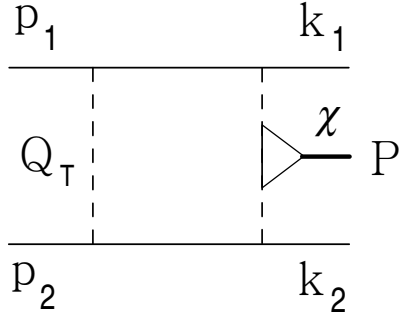}
\end{center}
\caption{Putting the inner quark lines ''on-shell'' is equivalent to
calculating the amplitude for $\chi$ meson production by double pomeron
exchange in the forward scattering limit $i.e.$ $t_{1}=t_{2}=0$.}%
\label{chi}%
\end{figure}

First, one calculates the (non-reggeized) production amplitude in the forward
direction $i.e.$ for vanishing transverse momenta of the produced $\chi$ meson
and of the final hadrons. In this case the sum of the diagrams of Fig.
\ref{chi-all} can be approximately replaced by the $s$-channel discontinuity
of the first one \cite{Bial-Land}, shown in Fig. \ref{chi}.

In the second step one introduces phenomenologically the effects of
reggeization, transverse momentum dependence and the gap survival so that the
full amplitude becomes:\footnote{For DPE central inclusive $\chi_{c}$ and
$\chi_{b}$ mesons production we can neglect the additional gap spoiling effect
$i.e.$ the Sudakov effect. It is justified by the relatively small masses of
the produced mesons.}%
\begin{equation}
\mathcal{M}=\mathcal{M}_{0}\left(  \frac{s}{s_{1}}\right)  ^{\alpha(t_{2}%
)-1}\left(  \frac{s}{s_{2}}\right)  ^{\alpha(t_{1})-1}F(t_{1})F(t_{2}%
)\exp\left(  \beta\left(  t_{1}+t_{2}\right)  \right)  S_{\text{gap}%
}.\label{M_all}%
\end{equation}
Here $\mathcal{M}_{0}$ is the amplitude in the forward scattering limit given
by the diagram shown in Fig. \ref{chi}. $\alpha\left(  t\right)
=1+\epsilon+\alpha^{\prime}t$ is the pomeron Regge trajectory with
$\epsilon\approx0.08,$ $\alpha^{\prime}=0.25$ GeV$^{-2}$. $s=(p_{1}+p_{2}%
)^{2}$, $s_{1}=(k_{1}+P)^{2},$ $s_{2}=(k_{2}+P)^{2},$ $t_{1}=(p_{1}-k_{1}%
)^{2}$ , $t_{2}=(p_{2}-k_{2})^{2}$ where $p_{1}$, $p_{2}$, $k_{1}$, $k_{2}$
and $P$ are defined in Fig. \ref{chi}. $F\left(  t\right)  $ = $\exp(\lambda
t)$ is the nucleon form-factor with $\lambda=$ $2$ GeV$^{-2}$. The
phenomenological factor $\exp\left(  \beta\left(  t_{1}+t_{2}\right)  \right)
$ with $\beta$ $=$ $1$ GeV$^{-2}$ takes into account the effect of the
momentum transfer dependence of the non-perturbative gluon propagator given
by ($p^{2}$ is the Lorentz square of the momentum carried by the non-perturbative
gluon):\footnote{As was stated in Ref. \cite{Bial-Land} there is no reason to
believe that the true form of $D$ is as simple as this. We hope it is not a
serious objection to our model.}
\begin{equation}
D\left(  p^{2}\right)  =D_{0}\exp\left(  p^{2}/\tau^{2}\right)
,\label{propag}%
\end{equation}
with $\tau=1$ GeV and $D_{0}G^{2}\tau=30$ GeV$^{-1}$ \cite{Bial-Land} where
$G$ is the scale of the process independent non-perturbative quark gluon coupling.

The factor $S_{\text{gap}}$ takes the gap survival effect into account $i.e.$
the probability ($S_{\text{gap}}^{2}$) of the gaps not to be populated by
secondaries produced in the soft rescattering. It is not a universal number
but it depends on the initial energy and the particular final state.
Theoretical predictions of the gap survival factor $S_{\text{gap}}^{2}$ can be
found in Ref. \cite{S2-theory}. In our calculations, following
\cite{Bzdak-jj,Bzdak-H}, we take for the Tevatron (LHC) energy $S_{\text{gap}%
}^{2}/\left(  G^{2}/4\pi\right)  ^{2}=0.6$ $(0.25)$.

Before we present the details of our calculation one point must be emphasized. 
Here we calculate in a model dependent way the non-perturbative part of the whole 
amplitude. The contribution coming from small distances, being another part of the 
whole amplitude, is beyond our approach. However, for the case of $\chi$ meson 
production the main part of the cross section may has a non-perturbative origin.                      

Following the calculation presented in Ref. \cite{Bial-Land} we find
$\mathcal{M}_{0}$ for colliding hadrons\footnote{The calculation of
$\mathcal{M}_{0}$ for colliding hadrons is performed in two steps. First, we
calculate the diagram presented in Fig. \ref{chi} for quark-quark scattering.
Next, we multiply by a factor $3^{2}$ to take the presence of three quarks in
each (anti)proton into account.} to be in the form:\footnote{This formula is
only valid in the limit of $\delta_{1,2}<<1$ where $\delta_{1,2}%
\equiv1-k_{1,2}/p_{1,2}$, and for small momentum transfer between initial and
final quarks.}%
\begin{equation}
\mathcal{M}_{0}=\frac{2G^{4}D_{0}^{3}}{\pi^{2}}\int d^{2}\vec{Q}_{\intercal
}\text{ }p_{1}^{\lambda}V_{\lambda\nu}^{J}p_{2}^{\nu}\exp(-3\vec{Q}%
_{\intercal}^{2}/\tau^{2}). \label{M_o}%
\end{equation}
Here $Q_{\intercal}$ is the transverse momentum carried by each of the three
gluons. $V_{\lambda\nu}^{J}$ is the $gg\rightarrow\chi^{J}$ vertex depending
on the polarization $J$ of the $\chi^{J}$ meson state. It was shown
\cite{Feng} that the DPE contribution to $\chi^{1}$ and $\chi^{2}$ production
in the forward scattering limit is vanishing (either perturbative or
non-perturbative two gluon exchange models).

It turns out that for $J=0$ we have the following simple result
\cite{Bial-Land,KMR-chi1,KMR-chi2}:
\begin{equation}
p_{1}^{\lambda}V_{\lambda\nu}^{0}p_{2}^{\nu}=\frac{s\vec{Q}_{\intercal}^{2}%
}{2M_{\chi^{0}}^{2}}A, \label{p_1Vp_2}%
\end{equation}
where $A$ is expressed by the mass $M_{\chi^{0}}$and the width $\Gamma
_{\chi^{0}}$of the $\chi^{0}$ meson through the relation:%
\begin{equation}
\Gamma_{\chi^{0}}=\frac{A^{2}}{2!4\pi M_{\chi^{0}}}. \label{A^2}%
\end{equation}

Substituting (\ref{M_o}) with (\ref{p_1Vp_2}) and (\ref{A^2}) to the full
amplitude (\ref{M_all}) and then performing the appropriate calculations
\cite{Bzdak-jj} we find the differential cross section $d\sigma/dy$ for DPE
central inclusive $\chi^{0}$ meson production to be in the form:
\begin{equation}
\frac{d\sigma}{dy}=\left(  \frac{s}{M_{\chi^{0}}^{2}}\right)  ^{2\epsilon
}\frac{CR^{2}}{\left(  (\lambda+\beta)/\alpha^{\prime}+\ln\left[  \sqrt
{s}/M_{\chi^{0}}\right]  \right)  ^{2}-y^{2}}. \label{dsigm/dy}%
\end{equation}
Here $y$, $\ln(M_{\chi^{0}}/(\delta_{2}^{\max}\sqrt{s}))\leqslant
y\leqslant\ln(\delta_{1}^{\max}\sqrt{s}/M_{\chi^{0}})$, is a rapidity of the
produced $\chi^{0}$ where $\delta_{1}$ and $\delta_{2}$ are defined as
$\delta_{1,2}\equiv1-k_{1,2}/p_{1,2}$ ($k_{1},$ $k_{2},$ $p_{1},$ $p_{2},$ are
defined in Fig. \ref{chi}). In the following we take $\delta_{1}^{\max}%
=\delta_{2}^{\max}=\delta=0.1$.

The factors $C$ and $R$ are defined as:%
\begin{equation}
C=\frac{18}{(12\pi)^{6}}\frac{\Gamma_{\chi^{0}}}{M_{\chi^{0}}^{3}}%
\frac{1}{\alpha^{\prime2}}\left(  D_{0}G^{2}\tau\right)  ^{6}\tau
^{2}\frac{S_{\text{gap}}^{2}}{(G^{2}/4\pi)^{2}},\label{constant}%
\end{equation}%
\begin{equation}
R=9\int d\vec{Q}_{\intercal}^{2}\text{ }\vec{Q}_{\intercal}^{2}\exp(-3\vec
{Q}_{\intercal}^{2})=1.\label{R}%
\end{equation}
In the expression (\ref{dsigm/dy}) $R^{2}$ reflects the structure of the loop
integral and is shown explicitly for the reason which will become clear in the following.

Performing integration over the whole range of rapidity we obtain the
following result for the central inclusive total cross section:%
\begin{equation}
\sigma=\left(  \frac{s}{M_{\chi^{0}}^{2}}\right)  ^{2\epsilon}\frac{CR^{2}%
}{(\lambda+\beta)/\alpha^{\prime}+\ln\left[  \sqrt{s}/M_{\chi^{0}}\right]
}\ln\left(  \frac{(\lambda+\beta)/\alpha^{\prime}+\ln(\delta s/M_{\chi^{0}%
}^{2})}{(\lambda+\beta)/\alpha^{\prime}-\ln\delta}\right)  .\label{total}%
\end{equation}
This completes the calculation of the cross section.

\section{Sudakov Factor}

The calculation presented in the previous section, based on the original
Bialas-Landshoff model, is a central inclusive one, $i.e.$ the QCD radiation
accompanying the produced object is allowed.\footnote{Let us notice that the
central inclusive process contains exclusive process. It implies that the
relation $\sigma_{\text{c.incl.}}>\sigma_{\text{excl.}}$ holds.} Thus, to
describe the exclusive processes where the central object is produced alone
one has to forbid this radiation. To this end we shall include the Sudakov
survival factor $T(Q_{\intercal},\mu)$ \cite{Khoze-all} inside the loop
integral over $Q_{\intercal}$ (\ref{R}). The Sudakov factor $T(Q_{\intercal
},\mu)$ is the survival probability that a gluon with transverse momentum
$Q_{\intercal}$ remains untouched in the evolution up to the hard scale
$\mu=M/2$.

Naturally a question of internal consistency arises. Namely, the Sudakov
factor uses perturbative gluons whilst in our calculations of the Born
amplitude (\ref{M_o}) we used non-perturbative gluons. We hope however that
taking the Sudakov factor in the loop integral into account we obtain an
approximate insight into exclusive processes. Moreover, it is shown
\cite{Bzdak-H} that such approach leads to the satisfactory description of the
DPE exclusive dijet production cross sections \cite{CDF-1,CDF-2}.

We take the function $T(Q_{\intercal},\mu)$ to be in the form \cite{Khoze-all}%
:%
\begin{equation}
T(Q_{\intercal},\mu)=\exp\left(  -\int_{\vec{Q}_{\intercal}^{2}}^{\mu^{2}%
}\frac{\alpha_{s}\left(  \vec{k}_{\intercal}^{2}\right)  }{2\pi}\frac{d\vec
{k}_{\intercal}^{2}}{\vec{k}_{\intercal}^{2}}\int_{0}^{1-\Delta}\left[
zP_{gg}\left(  z\right)  +\sum\limits_{q}P_{qg}(z)\right]  dz\right)  .
\label{T-def}%
\end{equation}
Here $\Delta=\left|  k_{\intercal}\right|  /\left(  \mu+\left|  k_{\intercal
}\right|  \right)  $, $P_{gg}\left(  z\right)  $ and $P_{qg}(z)$ (we take
$q=u,d,s,\bar{u},\bar{d},\bar{s}$) are the GLAP spitting functions.
$\alpha_{s}$ is the strong coupling constant.\footnote{In the following we take
$\alpha_{s}$ at one loop accuracy $i.e.$ $\alpha_{s}\left(  q^{2}\right)
=(4\pi/\beta_{0})\left(  1/\ln\left(  q^{2}/\Lambda^{2}\right)  \right)  $
with $\beta_{0}=9$ and $\Lambda=200$ MeV. Below $q=0.8$ GeV we freeze $\alpha_{s}$
to be $0.5$.} Taking into account the leading-order contributions \cite{Splitting}
to the GLAP splitting functions:%
\begin{align}
P_{gg}\left(  z\right)   &  =6\left[  z/(1-z)+(1-z)/z+z\left(  1-z\right)
+\delta\left(  1-z\right)  \left(  11/2-n_{f}/3\right)  \right]  ,\nonumber\\
P_{qg}\left(  z\right)   &  =\left[  z^{2}+\left(  1-z\right)  ^{2}\right]
/2, \label{Split}%
\end{align}
we obtain:%
\begin{align}
\int_{0}^{1-\Delta}zP_{gg}\left(  z\right)  dz  &  =-11/2+12\Delta-9\Delta
^{2}+4\Delta^{3}-3\Delta^{4}/2-6\ln\Delta,\nonumber\\
\int_{0}^{1-\Delta}P_{qg}\left(  z\right)  dz  &  =1/3-\Delta/2+\Delta
^{2}/2-\Delta^{3}/3.
\end{align}

Now to describe the exclusive processes we use the formulas (\ref{dsigm/dy}%
,\ref{total}) with $R^{2}=1$ replaced by $\tilde{R}^{2}\left(  \mu\right)  $
where $\tilde{R}\left(  \mu\right)  $ is defined as:\footnote{Notice that
$\mu>1.5$ GeV is required so that $9\int_{0}^{\mu^{2}}d\vec
{Q}_{\intercal}^{2}\,\vec{Q}_{\intercal}^{2}\exp(-3\vec{Q}_{\intercal}^{2}%
)=1$.}%
\begin{equation}
\tilde{R}\left(  \mu\right)  =9\int_{0}^{\mu^{2}}d\vec{Q}%
_{\intercal}^{2}\,\vec{Q}_{\intercal}^{2}\exp\left(  -3\vec{Q}_{\intercal}%
^{2}\right)  T\left(  Q_{\intercal},\mu\right)  . \label{R-tylda}%
\end{equation}

\section{Exclusive $\chi$ meson production. The CDF result}

Now we are ready to give our predictions for DPE exclusive $\chi_{c}^{0}$ and
$\chi_{b}^{0}$ mesons production at the Tevatron and the LHC energies. The
masses and widths of the $\chi_{c}^{0}$ and $\chi_{b}^{0}$ are taken as
follows: $M_{\chi_{c}^{0}}=3.4$ GeV, $M_{\chi_{b}^{0}}=9.8$ GeV, $\Gamma
_{\chi_{c}^{0}}=15$ MeV, $\Gamma_{\chi_{b}^{0}}=2.15$ MeV. Taking into account
(\ref{R-tylda}) we find the effective Sudakov suppression of the cross section
to be ($\mu=M/2$):
\begin{align}
\tilde{R}^{2}(\chi_{c}^{0})  &  =0.3,\nonumber\\
\tilde{R}^{2}(\chi_{b}^{0})  &  =0.07.
\end{align}

In Table \ref{chi_results} the predictions for the Tevatron and the LHC
energies are shown. As was discussed earlier we take $\delta=0.1$ and assume
$S_{\text{gap}}^{2}/\left(  G^{2}/4\pi\right)  ^{2}$ to be $0.6$ $(0.25)$ for
the Tevatron (LHC) energy.\begin{table}[h]
\begin{center}%
\begin{tabular}
[c]{|c|c|c|}\hline\hline
$\sqrt{s}$ & $d\sigma/dy(y=0)$ & $\sigma$\\\hline
$\chi_{c}^{0}%
\begin{array}
[c]{c}%
2\text{\ \thinspace TeV}\\
14\text{ TeV}%
\end{array}
$ & $%
\begin{array}
[c]{c}%
45\text{ nb}\\
30\text{ nb}%
\end{array}
$ & $%
\begin{array}
[c]{c}%
370\text{ nb}\\
350\text{ nb}%
\end{array}
$\\\hline
$\chi_{b}^{0}%
\begin{array}
[c]{c}%
2\text{\ \thinspace TeV}\\
14\text{ TeV}%
\end{array}
$ & $%
\begin{array}
[c]{c}%
0.05\text{ nb}\\
0.03\text{ nb}%
\end{array}
$ & $%
\begin{array}
[c]{c}%
0.3\text{ nb}\\
0.3\text{ nb}%
\end{array}
$\\\hline
\end{tabular}
\end{center}
\caption{Our results for DPE exclusive $\chi_{c}^{0},$ $\chi_{b}^{0}$ mesons
production cross section for the Tevatron and the LHC energies.}%
\label{chi_results}%
\end{table}

As can be seen from Table \ref{chi_results} the obtained results for the LHC
energy are comparable or even smaller than those for the Tevatron energy. It
is due to the faster decrease of the rapidity gap survival factor
$S_{\text{gap}}^{2}$ with increasing energy than the increase of our ''soft''
energy dependence $s^{2\epsilon}$. It should be noted, however, that the $s$
dependence is the most serious uncertainty of our approach and the results
presented in Table \ref{chi_results} should be regarded only as an order of
magnitude estimates. Bearing in mind this uncertainty our results seem to be 
comparable with those obtained in Refs. \cite{KMR-chi1,KMR-chi2,Feng}.

At this point one comment is to be in order. At first sight it seems that the calculation 
presented in this Letter based on the model with the effective gluon propagator of the 
exponential form $D_{0}\exp(-\vec{Q}_{\intercal}^{2})$ can be only used to estimate the 
contribution coming from the relatively low $Q_{\intercal}$. However, it is not the case. 
The reason is following. As was mentioned before we take for the Tevatron energy 
$S_{\text{gap}}^{2}/\left(  G^{2}/4\pi\right)  ^{2}=0.6$ \cite{Bzdak-jj,Bzdak-H}. If we 
assume the gap survival factor $S_{\text{gap}}^{2}$ to be $0.05$ \cite{KMR-chi2,S2-theory} we obtain 
the non-perturbative coupling $G^{2}/4\pi$ to be $0.3$. From the magnitude of the total cross 
section we conclude that $D_{0}G^{2}=30$ GeV$^{-2}$ \cite{Bial-Land} what allows us to extract 
$D_{0}=8$ GeV$^{-2}$.  
Now if we compare the numerical values of our non-perturbative gluon propagator 
$8\exp(-\vec{Q}_{\intercal}^{2})$ with the perturbative one $1/\vec{Q}_{\intercal}^{2}$ we 
observe that $8\exp(-\vec{Q}_{\intercal}^{2})>1/\vec{Q}_{\intercal}^{2}$ for 
$0.15<\vec{Q}_{\intercal}^{2}<3.3$ GeV$^{2}$. It means that the results presented in Table 
\ref{chi_results} include contribution coming not only from the non-perturbative region,
say $\vec{Q}_{\intercal}^{2}<1$ GeV$^{2}$, but also from the quasi-perturbative region 
$1<\vec{Q}_{\intercal}^{2}<3.3$ GeV$^{2}$. Moreover, we have checked that the contribution 
coming from the perturbative region $\vec{Q}_{\intercal}^{2}>3.3$ GeV$^{2}$ with the 
non-perturbative gluon propagators replaced by perturbative ones is negligible.

At the end, let us notice that the result $d\sigma/dy(y=0)=45$ nb for DPE
exclusive $\chi_{c}^{0}$ production is consistent with the preliminary CDF Run
II upper limit \cite{Koji}, see Table \ref{CDFresult}. Indeed, if we integrate the 
differential cross section (\ref{dsigm/dy}) over the rapidity range $\left|  y\right|  <0.6$
\cite{Koji} we obtain the value\footnote{Since the differential cross section
(\ref{dsigm/dy}) weekly depends on the rapidity of the produced meson it is
enough to multiply $d\sigma/dy(y=0)=45$ nb by a factor $1.2$.} $55$ nb to be
compared with the CDF upper limit $80$ $\pm$ $30$(stat) $\pm$ $70$(syst) nb
\cite{Koji}.\begin{table}[h]
\begin{center}%
\begin{tabular}
[c]{|c|c|c|}\hline\hline
$%
\begin{array}
[c]{c}%
\sqrt{s}=2\text{ TeV}\\
\left|  y\right|  <0.6
\end{array}
$ & CDF upper limit & Model\\\hline
$\chi_{c}^{0}$ & $80\pm100$ nb & $55$ nb\\\hline
\end{tabular}
\end{center}
\caption{Comparison of the CDF upper limit for DPE exclusive $\chi_{c}^{0}$
meson production with the result obtained in the presented model. The
satisfactory consistency is observed.}%
\label{CDFresult}%
\end{table}

\section{Conclusions}

In conclusion, in this paper we investigated the central inclusive and
exclusive DPE process of $\chi_{c}^{0}$, $\chi_{b}^{0}$ mesons production in
\textit{pp }(\textit{p\={p}}) collisions. We observed that the relation
$\sigma_{\text{excl.}}/\sigma_{\text{c.incl.}}\approx0.3$ ($\chi_{c}^{0}$),
$0.07$ ($\chi_{b}^{0}$) between exclusive and central inclusive cross sections
are hold. The recently proposed exclusive extension of the Bialas-Landshoff
model was found to be consistent with the preliminary CDF upper limit on
double diffractive exclusive $\chi_{c}^{0}$ production cross section.

\bigskip

\bigskip

\textbf{Acknowledgement} I would like to thank Dr. Leszek Motyka for helpful 
discussions. This investigation was supported by the Polish State
Committee for Scientific Research (KBN) under grant 2 P03B 043 24.

\end{document}